\documentclass[conference]{IEEEtran}
\IEEEoverridecommandlockouts
\usepackage{cite}
\usepackage{amsmath,amssymb,amsfonts}
\usepackage{algorithmic}
\usepackage{graphicx}
\usepackage{textcomp}
\usepackage{xcolor}
\usepackage{calrsfs}
\usepackage{balance}
\usepackage{caption}
\usepackage{subcaption}
\usepackage[ruled, lined, linesnumbered, commentsnumbered, longend]{algorithm2e}

\def\BibTeX{{\rm B\kern-.05em{\sc i\kern-.025em b}\kern-.08em
    T\kern-.1667em\lower.7ex\hbox{E}\kern-.125emX}}

\begin{document}
\title{Wireless Link Scheduling via Interference-aware Symmetric Positive Definite Connectivity Manifolds \vspace{-0.5in}
}
\vspace{-0.2in}
\author{\IEEEauthorblockN{\large{Ahmed S. Ibrahim}} \vspace{0.05in}
\IEEEauthorblockA{Department of Electrical and Computer Engineering, 
Florida International University,
Miami, Florida, USA \\
aibrahim@fiu.edu}
\vspace{-0.4in}
}

\maketitle
\vspace{-1.4in}

\newcommand\mycommfont[1]{\small\ttfamily\textcolor{blue}{#1}}
\SetCommentSty{mycommfont}

\begin{abstract}
In this paper, we investigate the fundamental problem of wireless link scheduling in device-to-device (D2D) networks, through the lens of Riemannian geometry. Our goal is to find a novel metric to characterize interference among D2D pairs, which can pave the way towards efficient and fast scheduling algorithms. Towards achieving this goal, we first model the connectivity pattern of each D2D pair, including its interference links, as a positively-shifted Laplacian matrix, which is a symmetric positive definite (SPD) one. Noting that SPD matrices constitute a non-Euclidean manifold, we represent each of the D2D pairs as a point on the SPD (i.e., conic) manifold, which is analyzed via Riemannian geometry. Accordingly we employ Riemannian metrics (e.g., Log-Euclidean metric ``LEM''), which are suitable measures of distances on manifolds, to characterize the interference among D2D points on the SPD manifold. 

To validate the effectiveness of the proposed \emph{LEM-based interference measure}, we propose a sequential link selection algorithm that schedules D2D pairs in a descending order of their signal-to-noise ratio (SNR), while keeping their LEM distances towards the already-scheduled pairs on the Riemannian manifold to be greater than a certain LEM threshold. Such LEM-based condition is equivalent to limiting the interference from potential D2D pairs to be below certain threshold. We show that the proposed LEM-based scheduling algorithm achieves sum rate of more than $86\%$ of state-of-the-art ones (e.g., FPLinQ~\cite{2017_FPLinQ_Yu}), while only requiring spatial locations of D2D pairs, as opposed to requiring full channel state information (CSI). 
\end{abstract}

\vspace{0.05in}
\begin{IEEEkeywords}
Conic manifolds, Laplacian matrices, link scheduling, Log-Euclidean metric, Riemannian geometry, symmetric positive definite matrices. 
\end{IEEEkeywords}

\vspace{-0.00in}

\section{Introduction} \label{sec_intro}

Link scheduling in device-to-device (D2D) communications, with full frequency reuse, is one of the fundamental challenges in wireless networks. Such challenge can be extended to other scenarios including vehicle-to-vehicle (V2V) communications. Aiming to maximize the sum rate and with full knowledge of the channel state information (CSI), such challenge can be formulated as non-convex combinatorial optimization problem, which is an NP-hard one~\cite{2017_FPLinQ_Yu}. Multiple practical scheduling schemes have been considered in the past such as FlashLinQ~\cite{2013_FlashLinQ_Wu}, which follows a sequential link selection approach. Alternatively, FPLinQ~\cite{2017_FPLinQ_Yu} utilizes fractional programming approach to iteratively solve maximum-rate optimization problem within a finite number of iterations. 

Towards the sole utilization of spatial locations of D2D pairs, as opposed to instantaneous CSI, deep neural networks (DNN) have been recently employed for wireless link scheduling in~\cite{2019_DNN_Scheduling_Yu}. As hundreds of thousands training samples (e.g., $800,000$ in~\cite{2019_DNN_Scheduling_Yu}) were required for such DNN, other DNN-based schemes have been pursued. For example, a graph embedding approach was utilized in the DNN-based solution of~\cite{2019_Li_Graph_Scheduling} to reduce the number of training layouts to $500$, while still requiring no instantaneous CSI. In doing so, the graph embedding approach in~\cite{2019_Li_Graph_Scheduling} modeled each D2D pair as a vertex in a graph, while modeling the interference between each two D2D pairs as an edge between their corresponding vertices. Having such innovative modeling of interference among D2D pairs can result in more efficient scheduling algorithms, and this is the main \emph{motivation} of this work.


This paper is particularly motivated by the need to find some unexplored characteristics in wireless networks (e.g., underlying non-Euclidean structures), which can be utilized to efficiently represent interference among D2D pairs, and hence can lead to more efficient scheduling algorithms. To that end, we first model each D2D pair by its connectivity graph, which includes its desired direct link along with its interference-related ones. A Laplacian matrix for each connectivity graph can be consequently computed via its incidence and weight matrices. Regularizing (e.g., positively-shifting) such Laplacian matrices results in symmetric positive definite (SPD) ones. We note that such SPD matrices can be represented on the interior of convex cones, or simply \emph{conic} manifolds (i.e., curved surfaces)~\cite{2015_Sra_Cone_SPD}. Therefore, each D2D pair's interference network can be represented as a point on conic manifolds, which follow non-Euclidean (i.e., non-flat) geometry~\cite{2014_Sra_SPD_NonEuc_Riem}. 

Conic or SPD structures are special class of \emph{Riemannian manifolds}~\cite{2018_Lee_Book_RiemManifolds}, which are equipped with \emph{Riemannian metrics} (i.e., distance functions) and are studied using \emph{Riemannian geometry}~\cite{2019_Intro_RiemGeometry}. Riemannian metrics, such as Log-Euclidean metric (LEM)~\cite{2006_LEM_Arsigny}, can be utilized to measure the distances among the D2D points on the SPD manifold. Intuitively, the larger the LEM distance between two D2D points is, the less their underlying interference graphs have in common, and hence the less interference they cause to each other. Therefore, we propose to employ LEM to measure interference among D2D representations on the Riemannian manifold, and this is the \emph{first contribution} of this paper. 

\begin{figure}[htbp]
	\vspace{-0.0in}
	\centerline{
		\includegraphics[width=0.45\textwidth]{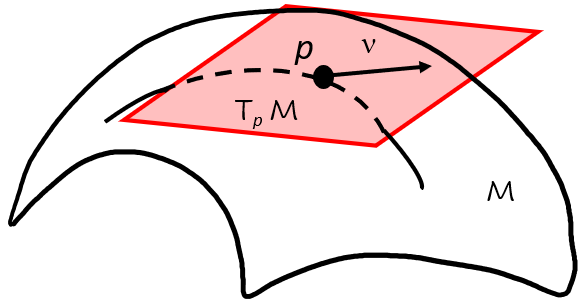}} \vspace{-0.05in}
	\caption{\small Non-Euclidean manifold $M$ and its $2$-dimensional tangent plane $T_p\,M$ at point $p$, which is a collection of all tangent vectors $\nu$.}
	\label{fig_manifold}
	\vspace{-0.2in}
\end{figure}

Aiming to validate the effectiveness of the proposed \emph{LEM-based interference measure}, we propose a link scheduling algorithm, which follows the general approach of sequential link selection. The proposed algorithm is the \emph{second contribution} of this paper and its novelty, as opposed to other sequential link selection ones (e.g., FlashLinQ~\cite{2013_FlashLinQ_Wu}), lies in characterizing the interference impact of any candidate D2D pair as the minimum LEM distance between itself and each of the already-scheduled ones. Consequently, the proposed algorithm schedules links in a descending order of their signal-to-noise ratio (SNR), as long as their LEM-based interference measure is greater than a certain threshold. Such threshold comparison is equivalent to guaranteeing the potential interference to be less than a given threshold, and it is set in an adaptive manner depending on each network layout. 

We note that Riemannian geometry has been recently considered in designing beamforming vectors (e.g.,~\cite{2019_Fan_Riem_MassiveMIMO, 2017_Chen_Riem_MUI}). 
Moreover, non-Euclidean methods have been utilized in codebook design such as \emph{non-conic} Riemannian manifolds (i.e, unitary, fixed rank) as in~\cite{2019_Schober_BF_Riem,2016_Letaief_Riem_mmWave_Precoding}. While these research works present novel non-Euclidean geometric perspectives of beamforming and codebook designs, they have not been utilized in link scheduling or to exploit the SPD structure of interference networks, as proposed in this paper.

The rest of this paper is organized as follows. The K-user interference channels and their mapping to Riemannian manifolds is presented in Section~\ref{sec_mod}. Section~\ref{sec_problem} introduces the problem formulation of maximizing the sum rate as well as its proposed equivalent on the Riemannian manifold. Section~\ref{solution} presents the proposed LEM-based sequential link selection algorithm. The simulation results are presented in Section~\ref{sim_results}. Finally, Section~\ref{conc} concludes the paper and discusses the way forward to build on the findings of this paper.

\section{System Model} \label{sec_mod}
In this section, we present brief preliminaries on Riemannian geometry including manifolds and metrics. Then, we introduce introduce the system model starting with modeling each D2D pair as an Euclidean-based graph then as a point on non-Euclidean Riemannian manifold.

\begin{figure}[htbp]
	\vspace{-0.0in}
	\centerline{
		\includegraphics[width=0.2\textwidth]{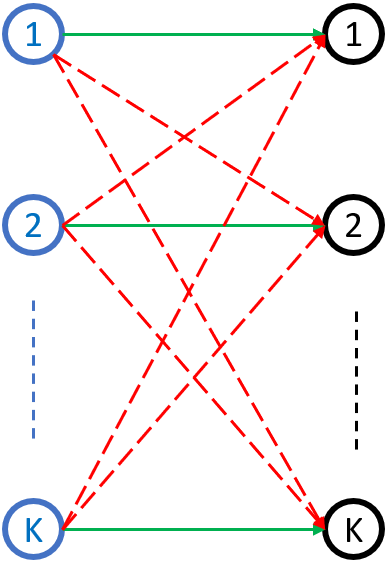}} \vspace{-0.05in}
	\caption{\small K-user interference channel.}
	\label{fig_k-user}
	\vspace{-0.2in}
\end{figure}
\subsection{Riemannian Geometry}
Topological manifolds are spaces that locally resemble the $n$ dimensional real coordinate space $\mathbb{R}^n$. Differential manifolds~\cite{2016_DiffGeom_Carmo} are topological ones with \emph{smooth} changes of coordinates (maps from $\mathbb{R}^n$ to $\mathbb{R}^n$)~\cite{2014_Godinho_Book_RiemGeom}. Fig.~\ref{fig_manifold} shows tangent space $T_p\, M$ of a differential manifold $M$, at a point $p \in M$, which is a vector space of all vectors $\nu$ that are tangent to manifold $M$ at point $p$. \emph{Riemannian} geometry is the study of Riemannian manifolds $(M,g)$~\cite{2018_Lee_Book_RiemManifolds, 2019_Intro_RiemGeometry}, which are differential manifolds $M$ with metric $g$. Riemannian metric is a smooth inner product on tangent spaces of a manifold, and it measures length of geodesics (i.e., curves) on the manifold. SPD matrices lie on the interior of convex cones (i.e., conic manifolds or curved  surfaces)~\cite{2015_Sra_Cone_SPD}, which are special class of Riemannian manifolds~\cite{2018_Lee_Book_RiemManifolds}. Riemannian metrics for SPD matrices, which must respect the non-Euclidean geometry of SPD matrices, have been previously proposed such as LEM~\cite{2006_LEM_Arsigny} and Affine-Invariant Metric (AIM)~\cite{Pennec2006_AIM}.

\subsection{Modeling D2D Pairs as Multiple Graphs}
Fig.~\ref{fig_k-user} illustrates the K-user interference channel, which is the underlying model for D2D or V2V communications. It consists of $K$ users or D2D pairs, and each pair has one transmitter (in blue) and one receiver (in black). As each transmitter communicates with its intended receiver and with full frequency reuse, it causes interference to other receivers. 

The shown network of $K$ D2D pairs in Fig.~\ref{fig_k-user} includes $n = 2\, K$ nodes, which are the $K$ transmitters and $K$ receivers. Two D2D pairs $\{k_1, k_2\}$, where $\{k_1, k_2\}=1,2,\cdots,K$ and $k_1 \neq k_2$, can be represented as a \emph{weighted} and \emph{directed} finite graph $G_{k_1, k_2} (V,E_{k_1, k_2})$, where $V=\{v_1, v_2, \cdots, v_n\}$ is the set of all $n$ nodes and $E_{k_1, k_2}$ is the set of all $m_{k_1, k_2}$ edges (links) that are connecting transmitters of $\{k_1, k_2\}$ pairs to their intended receivers along with the interference links affecting the receivers of the $\{k_1, k_2\}$ pairs. The \emph{weight} of interfering edges is set to the Euclidean distance between its endpoints. Moreover, the weight of a desired edge, between a transmitter and its intended receiver, is normalized to $1$.

For a \emph{directed} edge $l$, $1 \leq l \leq m_{k_1, k_2}$, from node $v_i$ to node $v_j$, $\{v_i,v_j\}\in V$, we define an edge vector ${\mathbf{a_{l}}} \in \mathbb{R}^n$, where the $i$-th and $j$-th elements are given by $a_{l,i} = 1$ and $a_{l,j} = -1$, respectively, and the rest is zero. The incidence matrix $\mathbf{A_{k_1, k_2}} \in {\mathbb{R}^{n \times m_{k_1, k_2}}}$ of the graph $G_{k_1, k_2}$ is the matrix with $l$-th column given by $\mathbf{a_{l}}$. The weight matrix $\mathbf{W_{k_1, k_2}} \in {\mathbb{R}^{m_{k_1, k_2} \times m_{k_1, k_2}}}$ is a diagonal matrix, whose $(l,l)$-th element is equal to the weight of the $l$-th edge. Finally, the \emph{Laplacian} matrix $\mathbf{L_{k_1, k_2}} \in \mathbb{R}^{n \times n}$ is defined as  
\begin{equation} 
\mathbf{L_{k_1, k_2}}  = \mathbf{A_{k_1, k_2}}\; \mathbf{W_{k_1, k_2}} \; \mathbf{A_{k_1, k_2}}^T \,,  
\label{L_matrix}
\end{equation}
where $\{k_1, k_2\}=1,2,\cdots,K$, $k_1 \neq k_2$, and $T$ denotes matrix transposition. 

A special case of the $G_{k_1, k_2}$ graph is the one with single direct-link only, denoted as $G_{k}$ for the $k$-th directed D2D pair, $k=1,2,\cdots,K$. It can be represented as a finite \emph{directed} graph $G_k(V,E_k)$, where $E_k$ includes only $1$ edge from the $k$-th transmitting node to the $k$-th receiving node, with normalized weight of $1$. The incidence matrix in this case $\mathbf{A_k} \in {\mathbb{R}^{n \times 1}}$ has $1$ and $-1$ at the positions of the $k$-th transmitting and receiving nodes, respectively, and the rest is zero. The Laplacian matrix in this case is $\mathbf{L_k} = \mathbf{A_k}\, \mathbf{A_k}^T$, as its weight matrix is $1$. 

\subsection{Modeling D2D Links over Riemannian Manifolds} \label{D2D_Manifold}
The Laplacian matrices in (\ref{L_matrix}) are positive semi-definite. Accordingly, a simple regularization step~\cite{2015_Riem_Brain_Class} can be implemented by adding a scaled identity matrix to the Laplacian matrix to produce a \emph{regularized} SPD Laplacian matrix as 
\begin{align}
 \mathbf{S_{k_1, k_2}} & =  \mathbf{A_{k_1, k_2}}\, \mathbf{W_{k_1, k_2}} \, \mathbf{A_{k_1, k_2}}^T + \gamma \, \mathbf{I} \,, \label{S_matrix_1} \\
 \mathbf{S_k} & =  \mathbf{A_k}\,\mathbf{A_k}^T+ \gamma \, \mathbf{I}  \,,
\label{S_matrix_2}
\end{align}
where $\mathbf{I}$ is the $n \times n$ identity matrix and $\gamma$ is an arbitrary small scalar (e.g., $\gamma=0.5$). 

The regularized Laplacian matrices, $\mathbf{S_{k_1, k_2}}$ or $\mathbf{S_k}$, can be represented on the interior of convex cones~\cite{2012_Sra_SPD_Riem, 2015_Sra_Cone_SPD}, which are special class of Riemannian manifolds~\cite{2018_Lee_Book_RiemManifolds}. Therefore, Riemannian metrics such as LEM~\cite{2006_LEM_Arsigny}, a.k.a., log-Frobenius distance, can be utilized to measure the distance between regularized Laplacian matrices that correspond to different D2D pairs. For example, the LEM distance between $\mathbf{S_{k_1, k_2}}$ and $\mathbf{S_{k_1}}$ can be computed as
\begin{equation}
	\mathcal{D}(\mathbf{S}_{k_1, k_2},\mathbf{S}_{k_1})= ||\log(\mathbf{S}_{k_1, k_2}) - \log(\mathbf{S}_{k_1})||_F^2 \;, 
	\label{eqn_LEM}
\end{equation}
where $||\,.\,||_F$ denotes the matrix Frobenius norm. 

The LEM distance in (\ref{eqn_LEM}) can be utilized in quantifying the interference the $k_2$-th D2D pair causes to the $k_1$-th D2D pair, if scheduled simultaneously. Therefore such LEM distance can be utilized in link scheduling decision, as will be formulated in the next section. 


\section{Problem Formulation}\label{sec_problem}
This section presents the conventional sum-rate problem formulation in K-user interference channel, then maps it to the novel formulation over SPD manifolds.

In the K-user interference channel, previously shown in Fig.~\ref{fig_k-user}, let $x_i \in \{0,1\}$ be a binary variable denoting if the $i$-th link, $i=1,2,\cdots,K$, will be scheduled or not. Let $\mathbf{\underline{x}}= [x_1, x_2, \cdots, x_K]^T$ be a $K \times 1$ vector that contains all these variables. We aim to find the optimum values of such binary variables, which maximize the summation of the individual information-theoretic rates over bandwidth $B$, as in
\begin{align} 
	\max_{\mathbf{\underline{x}}} & \;  \sum_{i=1}^{K} B \, \log_2 \Big(1 + \frac{p \, x_i \, |h_{ii}|^2 \, d_{ii}^{-\alpha}}{ \sum_{j\neq i} \;\; p \, x_j \, |h_{ij}|^2 \, d_{ij}^{-\alpha} + \sigma^2}\Big) \;, \nonumber \\
	&\text{s.t.} \;\; x_{i} \in \{0,1\}\;, \;  i = 1, 2, \cdots, K \,, 
	\label{opt_RA}
\end{align}
where $p$ is the transmission power, which is the same for all links. The $d_{i,j}$ and $h_{i,j}$ in (\ref{opt_RA}) are the Euclidean distance and fast-fading channel gain (i.e., CSI), respectively, between the $j$-th transmitter and $i$-th receiver. Finally in (\ref{opt_RA}), $\alpha$ is the path loss exponent and $\sigma^2$ is the noise variance. The optimization problem in (\ref{opt_RA}) is a challenging combinatorial NP-hard one as the optimal value for each binary variable $x_i$ depends on the choices of the other ones~\cite{2017_FPLinQ_Yu}.

It was proposed in Section~\ref{D2D_Manifold} that the LEM distance between two regularized Laplacian matrices, computed as in~(\ref{eqn_LEM}), can be employed to measure the interference between them. Therefore, we propose to decouple the signal and interference terms in~(\ref{opt_RA}) and utilize the LEM as the interference measure. More precisely, we reformulate the link scheduling problem in~(\ref{opt_RA}) to aim to maximize the received signal powers or SNRs, while limiting the LEM-based interference measure to be above certain threshold as
\begin{align}
	\vspace{-0.6cm}
	\max_{\mathbf{\underline{x}}}  \sum_{i=1}^{K} & x_i \, \frac{p \, |h_{ii}|^2 \, d_{ii}^{-\alpha}}{ \sigma^2}\;, \label{opt_RA_LEM_obj} \\
	\text{s.t.} \;\; & \min_{i,j} \mathcal{D}(\mathbf{L}_{i,j},\mathbf{L}_i) \geq \Delta \;, \; \forall i, j \in \mathcal{Z}  , \label{opt_RA_LEM_const} \\ & x_{i} \in \{0,1\}\;, \;  i = 1, 2, \cdots, K , \nonumber 
	\label{opt_RA_LEM} \vspace{-0.6cm}
\end{align}
where $\mathcal{Z}$ denotes the set of scheduled D2D pairs. 

The objective function in (\ref{opt_RA_LEM_obj}) represents the summation of SNRs of scheduled links. Increasing such objective function leads to higher sum-rate, which is the original objective function in~(\ref{opt_RA}). Furthermore, the constraint in~(\ref{opt_RA_LEM_const}) guarantees that adding a new D2D pair $j$ does not decrease the piece-wise LEM distances $\mathcal{D}(\mathbf{L}_{i,j},\mathbf{L}_i)$, computed towards the already-scheduled links $i \in \mathcal{Z}$, to be below certain threshold $\Delta$. In other words, the constraint in~(\ref{opt_RA_LEM_const}) forces the minimum piece-wise LEM distance across all scheduled links $i,j$ to be greater than a specific threshold $\Delta$. Together, the objective function and constraints of the proposed LEM-based problem formulation aim to select links with maximum SNR, and hence higher rate, while maintaining large LEM distance (or equivalently, minimum interference level) towards the already-scheduled links. 

\section{ Wireless Link Scheduling via Interference-aware Riemannian Manifolds} \label{solution}

\begin{algorithm}
	\SetKwFunction{isOddNumber}{isOddNumber}
	\SetKwInOut{KwIn}{Inputs}
	\SetKwInOut{KwOut}{Output}
	\SetKwInOut{KwInit}{Initialization}
	
	\KwIn{$\mathbf{A}_{k_1}$, $\mathbf{A_{k_1, k_2}}$, and $\mathbf{W_{k_1, k_2}}$, where $\{k_1, k_2\}=1, 2, \cdots,K$ and $k_1 \neq k_2$.}
	
	\KwOut{Set $\mathcal{Z}$ of scheduled (active) pairs.	}  
	
	\KwInit{$\mathcal{Z} = \emptyset$, $\mathcal{Z}^c = \emptyset$} 
		
	\textbf{Sort} $K$ D2D pairs in set $\mathcal{Y}$, following an ascending order in terms of their direct transmission distances. 
	
	\textbf{Choose} first scheduled link to be the first one in $\mathcal{Y}$, i.e.,  $\mathcal{Z}.append\big(\mathcal{Y}(1)\big)$.
	
	$\mathbf{S}_{\mathcal{Y}(1)} = \mathbf{A}_{\mathcal{Y}(1)} \, \mathbf{A}^T_{\mathcal{Y}(1)} + \gamma \, \mathbf{I} $
	
	\For{$k_2 \in \mathcal{Y}$, $k_2 \neq \mathcal{Y}(1)$}{	
		
		$\mathbf{S}_{\mathcal{Y}(1), k_2} =  \mathbf{A}_{\mathcal{Y}(1), k_2}\, \mathbf{W}_{\mathcal{Y}(1), k_2} \, \mathbf{A}^T_{\mathcal{Y}(1), k_2} + \gamma \, \mathbf{I}$	
		
		$\mathcal{D}(\mathbf{S}_{\mathcal{Y}(1), k_2},\mathbf{S}_{\mathcal{Y}(1)})= ||\log(\mathbf{S}_{\mathcal{Y}(1), k_2}) - \log(\mathbf{S}_{\mathcal{Y}(1)})||_F^2$ 
	}	
	\textbf{Set} LEM distance \emph{threshold} to be $\Delta = r \times \max_{k_2} \mathcal{D}(\mathbf{S}_{\mathcal{Y}(1), k_2},\mathbf{S}_{\mathcal{Y}(1)})$, where $r=0.8$. 
	
	\For{$k_2 \in \mathcal{Y}$, $k_2 \not\in (\mathcal{Z} \cup \mathcal{Z}^c$)}{	
		
		\For{$k_1 \in \mathcal{Z}$}{
			$\mathbf{S}_{k_1, k_2} =  \mathbf{A}_{k_1, k_2}\, \mathbf{W}_{k_1, k_2} \, \mathbf{A}^T_{k_1, k_2} + \gamma \, \mathbf{I}$
			
			$\mathcal{D}(\mathbf{S}_{k_1, k_2},\mathbf{S}_{k_1})= ||\log(\mathbf{S}_{k_1, k_2}) - \log(\mathbf{S}_{k_1})||_F^2$ 
			
			\eIf{$\min_{k_1} \mathcal{D}(\mathbf{S}_{k_1, k_2},\mathbf{S}_{k_1}) \geq \Delta$}{
				
				$\mathcal{Z}.append (k_2)$ 
				
				$\mathbf{S}_{k_2} = \mathbf{A}_{k_2} \, \mathbf{A}^T_{k_2} + \gamma \, \mathbf{I} $
			}{
				$\mathcal{Z}^c.append (k_2)$ 
			}
		}
	}
	\KwRet{$\mathcal{Z}$}
	\caption{LEM-based Sequential Link Selection.} \label{algorithm}
\end{algorithm}

In this section, we describe the proposed link scheduling scheme, which is summarized in Algorithm~\ref{algorithm}. The \emph{inputs} to the algorithm are incidence matrices for individual pairs $\mathbf{A}_{k_1}$ and double-pairs $\mathbf{A_{k_1, k_2}}$, along with the corresponding weight matrices $\mathbf{W_{k_1, k_2}}$, where $\{k_1, k_2\}=1, 2, \cdots,K$ and $k_1 \neq k_2$. The \emph{output} is the set $\mathcal{Z}$, which includes the scheduled pairs for transmissions, i.e., $\mathcal{Z}$ contains the indices $i$, where $x_i= 1$. At the \emph{initialization} phase, the set of scheduled links is initialized to the empty set $\mathcal{Z} = \emptyset$. Similarly, its complement set is also initialized to the empty set $\mathcal{Z}^c = \emptyset$, which will eventually include the list of inactive links.

Steps $1$ to $3$ in Algorithm~\ref{algorithm} aim to choose the first scheduled link. Since the proposed algorithm follows a sequential link selection approach, it sorts the links in a descending order in terms of their SNR, or alternatively, in an ascending order in terms of the length of their direct links. So, shorter D2D links come first before the longer ones in the set $\mathcal{Y}$ of ordered links. The first scheduled D2D pair will be the first one in the ordered list $\mathcal{Y}$, i.e., $\mathcal{Z}.append\big(\mathcal{Y}(1)\big)$. Then, the regularized Laplacian matrix of this scheduled D2D pair alone is computed as in~(\ref{S_matrix_2}), where $k_1 = \mathcal{Y}(1)$. 

Steps $4$ to $8$ in Algorithm~\ref{algorithm} aim to set the LEM-based interference threshold as follows. The regularized Laplacian matrix of this first scheduled D2D pair along with each other one is computed as in (\ref{S_matrix_1}), where $k_1 = \mathcal{Y}(1), k_2 \in \mathcal{Y}, k_2 \neq k_1$. The LEM distance between the first scheduled D2D pair and each of the other ones is computed according to (\ref{eqn_LEM}). The LEM threshold $\Delta$ is set in reference to the maximum LEM distance towards the first scheduled link, as $\Delta = r \times \max_{j} \mathcal{D}(\mathbf{L}_{i,j},\mathbf{L}_i)$, where $0<r<1$. It was found via simulations, as will be shown later in Section~\ref{sim_results}, that $r=0.8$ produce high sum-rate, irrespective of the number of D2D pairs.
 
\begin{table}[htbp]
	\caption{\small Network simulation parameters.}
	\vspace{-0.1in}	
	\begin{center}
		\begin{tabular}{|l|l|}
			\hline
			\textbf{Parameter}&\textbf{Value} \\ \hline
			Deployment area & $500$ m $\times$ $500$ m\\ \hline
			D2D distance ($R$) & $2$ m to $65$m \\ 	\hline
			Transmit power over activated link $p$& $40$ dBm\\ 	\hline
			Antenna height for transmitters and receivers & $1.5$ m\\ \hline
			Antenna gain for transmitters and receivers & $2.5$ dB\\ \hline
			Carrier frequency & $2.4$ GHz\\ 	\hline
			Noise spectral density  & $-169$ dBm/Hz\\ \hline
			Bandwidth $B$ & $5$ MHz\\ \hline
		\end{tabular}
		\label{table_sim_parameters}
	\end{center}
	\vspace{-0.1in}	
\end{table}
 
Steps $9$ to $20$ in Algorithm~\ref{algorithm} conducts two nested loops with the goal of scheduling a D2D pair only if its LEM distances towards each of the already-scheduled ones are all greater than the LEM threshold $\Delta$. More precisely, an outer loop with running index $k_2$ runs over the links in $\mathcal{Y}$, while an inner loop with running index $k_1$ runs over the scheduled links in set $\mathcal{Z}$. For a given potential D2D pair $k_2 \in \mathcal{Y}$, its double-pair regularized Laplacian matrices and LEM distances towards all scheduled links $k_1 \in \mathcal{Z}$ are computed according to (\ref{S_matrix_1}) and (\ref{eqn_LEM}), respectively. If the minimum value of such LEM distances is greater than the LEM threshold, (i.e.,$\min_{k_1} \mathcal{D}(\mathbf{S}_{k_1, k_2},\mathbf{S}_{k_1}) \geq \Delta$) then the potential D2D pair $k_2$ is scheduled (i.e., $\mathcal{Z}.append(k_2)$) and its single-link regularized Laplacian matrix is computed as in (\ref{S_matrix_2}) for use in the remaining inner loop iterations. Otherwise, it is added to the set of unscheduled D2D pairs, i.e.,  $\mathcal{Z}^c.append(k_2)$.

\section{Simulation Results} \label{sim_results}

In this section, we present simulation results of the proposed scheduling scheme and compare it against existing ones. We follow the same simulation setup as in~\cite{2017_FPLinQ_Yu, 2019_DNN_Scheduling_Yu, 2019_Li_Graph_Scheduling} for fair comparison. An area of $500$ m $\times$ $500$ m is considered, in which a variable number of $K$ D2D pairs are deployed. The transmitter is uniformly deployed in this area, while the receiver is deployed within a disk around the transmitter with radius that is uniformly distributed between $2$ m and $65$ m. We consider distance-based path-loss according to the ITU-1411 model, with no shadowing effect. The fast-fading channel is randomly generated. The rest of the simulation parameters are included in Table~\ref{table_sim_parameters}. 

\begin{table}[!t]
	\vspace{0.1in}	
	\caption{\small Simulated scheduling schemes.}
	\vspace{-0.1in}	
	\begin{center}
		\begin{tabular}{|c|c|} 
			\hline \textbf{Scheme} & CSI \\
			\hline LEM (proposed) & No \\
			\hline FPLinQ~\cite{2017_FPLinQ_Yu}  & Yes \\
			\hline Greedy  & Yes \\			
			\hline Strongest Link Only & Yes \\	
			\hline Random & No \\	
			\hline All Active  & No\\	
			\hline
		\end{tabular}
		\label{table_schemes}
	\end{center}
	\vspace{-0.3in}	
\end{table}

Table~\ref{table_schemes} lists different simulated schemes, for comparison purposes. The state-of-the-art FPLinQ~\cite{2017_FPLinQ_Yu} scheme decouples the signal and interference terms in (\ref{opt_RA}) and uses a subsequent coordinated ascent approach to iteratively find the optimal solution. The greedy algorithm follows a sequential link scheduling approach as it adds a new link only if it does not reduce the sum rates of the already-scheduled links. The strongest link only algorithm only selects the link with the maximum SNR. All of the aforementioned algorithms require CSI for scheduling. The random algorithm schedules links randomly and the all active algorithm activates all links. Both of the last two algorithms do not require CSI.


%

\begin{figure}[htbp]
	\vspace{-0.0in}
	\centerline{
		\includegraphics[width=0.5\textwidth]{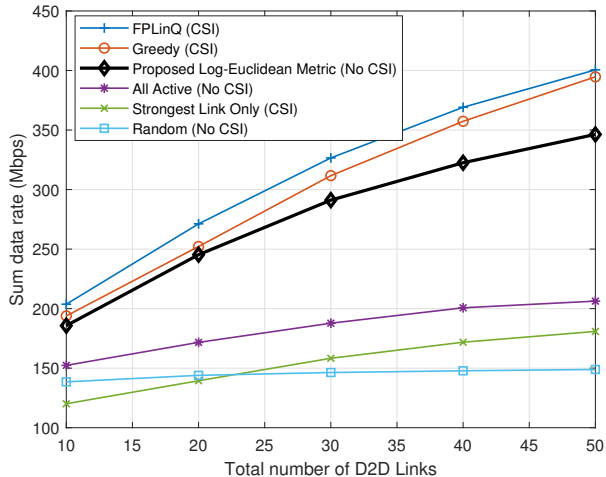}} \vspace{-0.05in}
	\caption{\small Achievable sum rate by various wireless link scheduling schemes.}
	\label{fig_rate}
	\vspace{-0.2in}
\end{figure}

Considering \emph{fast fading} scenario, Fig.~\ref{fig_rate} shows that achievable sum-rate by the different scheduling schemes. We note that the both FPLinQ and greedy algorithms make use of the full CSI to achieve high sum rate. While the proposed LEM-based sequential link selection algorithm does not utilize CSI, it achieves sum-rate of more than $86\%$ (at $K=50$ D2D pairs) compared to FPLinQ. It is also shown that the proposed LEM-based scheme significantly outperforms the strongest link only algorithm, as well as all other schemes that do not require CSI.

\begin{figure}[htbp]
	\vspace{-0.0in}
	\centerline{
		\includegraphics[width=0.5\textwidth]{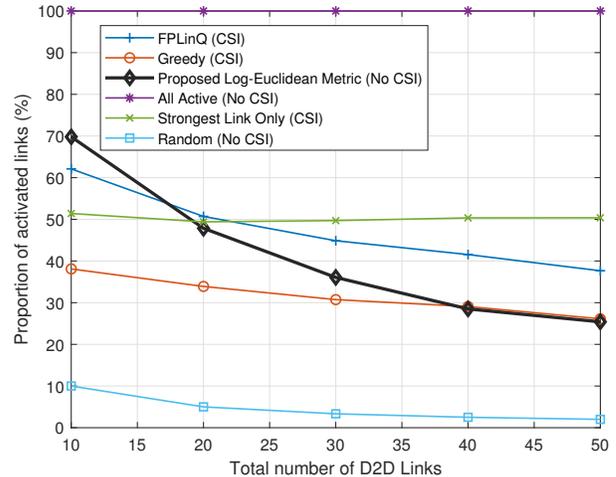}} \vspace{-0.05in}
	\caption{\small Activation (scheduling) ratio of D2D pairs, achieved by various wireless link scheduling schemes.}
	\label{fig_activation}
	\vspace{-0.2in}
\end{figure}

For better understanding of the behavior of the simulated scheduling schemes, Fig.~\ref{fig_activation} illustrates their activation or scheduling ratio. As shown, the proposed LEM-based scheduling algorithm have an activation ratio that is closer to the that of the FPLinQ, while not requiring full CSI. 

\section{Conclusion and Future Work} \label{conc}

In this paper, we have introduced a novel perspective to rethink the problem of wireless link scheduling in D2D communications through the lens of Riemannian geometry. First, we have shown that interference among D2D pairs can be measured on symmetric positive definite manifolds using Log-Euclidean Riemannian metric. Second, we have proposed a sequential link scheduling algorithm that decouples the desired signal maximization from limiting the LEM-based interference level. We have shown that the proposed LEM-based scheduling algorithm achieves more than $86\%$ of the maximum sum rate, while making use of the spatial locations of the D2D pairs alone instead of requiring full CSI.

This paper serves as a proof of concept that wireless link scheduling can be designed through mapping to non-Euclidean manifolds. Next, we will build on this paper by proposing conic geometric optimization approaches to find the scheduling decisions by searching over the SPD manifold. Furthermore, geometric machine learning models can be utilized to learn interference patterns over the SPD manifold. Both approaches aim to find low-complexity scheduling schemes.  

\vspace{0.0in}
\footnotesize \setlength{\baselineskip}{15pt}
\bibliographystyle{IEEEbib}
\bibliography{Ibrahim, Geometric_CAREER}

\end{document}